# SCIENTIFIC REPORTS

**OPEN**

# Period Increase and Amplitude Distribution of Kink Oscillation of Coronal Loop

W. Su[1,2,3,4], Y. Guo[3], R. Erdélyi[5,6], Z. J. Ning[2], M. D. Ding[3], X. Cheng[3] & B. L. Tan[4]

Coronal loops exist ubiquitously in the solar atmosphere. These loops puzzle astronomers over half a century. Solar magneto-seismology (SMS) provides a unique way to constrain the physical parameters of coronal loops. Here, we study the evolution of oscillations of a coronal loop observed by the Atmospheric Imaging Assembly (AIA). We measure geometric and physical parameters of the loop oscillations. In particular, we find that the mean period of the oscillations increased from 1048 to 1264 s during three oscillatory cycles. We employ the differential emission measure method and apply the tools of SMS. The evolution of densities inside and outside the loop is analyzed. We found that an increase of density inside the loop and decrease of the magnetic field strength along the loop are the main reasons for the increase in the period during the oscillations. Besides, we also found that the amplitude profile of the loop is different from a profile would it be a homogeneous loop. It is proposed that the distribution of magnetic strength along the loop rather than density stratification is responsible for this deviation. The variation in period and distribution of amplitude provide, in terms of SMS, a new and unprecedented insight into coronal loop diagnostics.

A single coronal loop may be visualized as a thin thread or elastic tube with a narrow but finite width compared to its length. These narrow loops can often be regarded as one-dimensional (1D) magnetic flux tubes because their lengths are much larger than their radii[1,2]. Generally, the magnetic pressure is much higher than the gas pressure in coronal loops, thus the plasma-$\beta$, which is the ratio of the gas to magnetic pressure, is low throughout the upper solar atmosphere where this loops reside mostly. Coronal loop oscillations were observed in EUV images by space telescopes in recent years[3–6], e.g. by the Transition Region and Coronal Explorer (TRACE), the Solar Terrestrial Relations Observatory (STEREO)[7], and the Solar Dynamics Observatory (SDO)[8]. Especially, observations of high temporal and spatial resolution by the Atmospheric Imaging Assembly (AIA)[9] on board SDO have revealed more and more details about coronal loop oscillations[10,11]. Most of the reported loop oscillations are identified as the asymmetric kink mode in the solar corona[12,13], but the symmetric sausage mode loop oscillations have also been observed and identified[14]. Occasionally, even Alfvén waves are reported, though often there is no consensus about these reports[15,16].

Coronal loop kink oscillations can be triggered directly by e.g. flares, EUV waves, or indirectly by, e.g., coupling through excitation of Alfvén waves at the loop footpoint[17–19]. The amplitude of the kink oscillations usually decays with time once oscillations are triggered[20,21]. Some decay-less kink oscillations have also been found[22,23]. Statistical studies reveal that the damping time is positively correlated with period, that is also consistent with theoretical predications[24–27]. A popular damping mechanism of loop oscillations is resonant absorption[28,29]. The cooling of the plasma has been proposed as another plausible and natural damping mechanism[30–32].

There is a way of combining the MHD seismological theories with observations to determine the physical parameters in the corona[33–35]. It has been demonstrated in theory that the Alfvén speed and even magnetic field strength can be obtained from measuring the periods of the kink oscillations[34]. Based on this approach, called solar magneto-seismology (SMS), studies have been carried out to infer the magnetic field of coronal loops from the observed periods of kink oscillations[5,36]. Besides the important information about magnetic field strength, another crucially needed input for constraining the theoretical approach, e.g. ratio of the inside and outside densities of a coronal loop can also be estimated by the SMS[35]. Alternatively, this density ratio can also be estimated by applying the differential emission measure (DEM) method[2]. Therefore, combining these two approaches may even enable us to study the fine (or sub-resolution) structure of coronal waveguides. Applying SMS together with magnetic field extrapolation techniques are often used to model the observed fundamental and harmonic modes of a kink oscillation[37].

[1]School of Physics and MOE Key Laboratory of Fundamental Physical Quantities Measurements, Huazhong University of Science and Technology, 1037 Luoyu Road, Wuhan, Hubei Province, 430074, China. [2]Key Laboratory of Dark Matter and Space Astronomy, Purple Mountain Observatory, Nanjing, 210008, China. [3]Key Laboratory for Modern Astronomy and Astrophysics (Nanjing University), Ministry of Education, Nanjing, 210023, China. [4]Key Laboratory of Solar Activity, National Astronomical Observatories, Chinese Academy of Sciences, Beijing, 100012, China. [5]Solar Physics and Space Plasma Research Centre (SP2RC), University of Sheffield, Hicks Building, Hounsfield Road, Sheffield, S3 7RH, UK. [6]Department of Astronomy, Eötvös Loránd University, Pázmány P. sétány 1/a, Budapest, H-1117, Hungary. Correspondence and requests for materials should be addressed to W.S. (email: suw12@hust.edu.cn) or to Z.J.N. (email: ningzongjun@pmo.ac.cn)

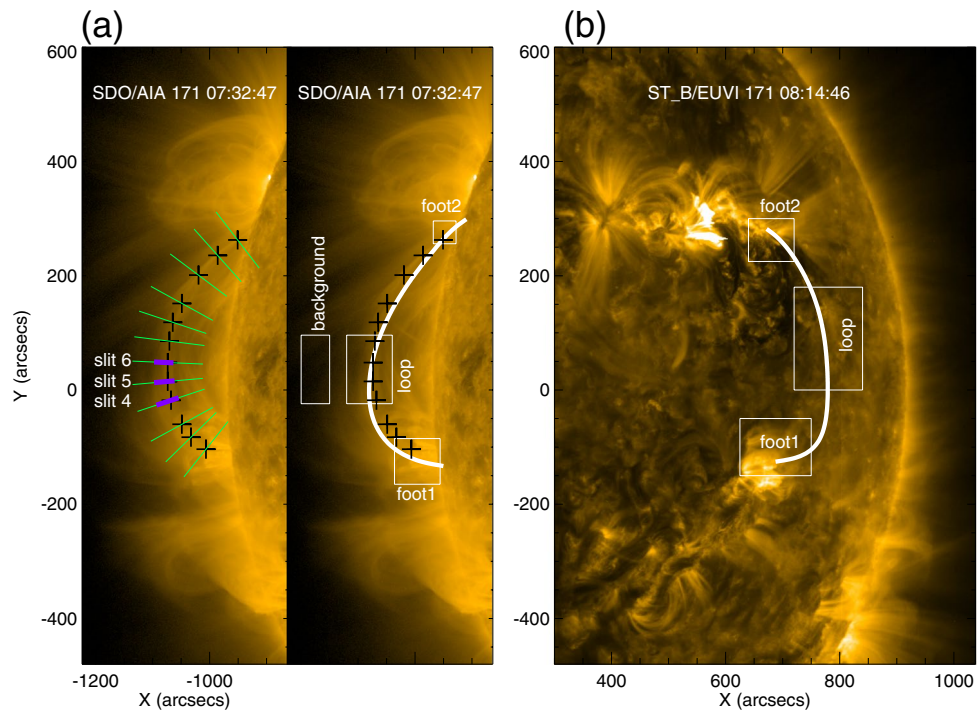

**Figure 1.** An overview of the coronal loop analyzed in this paper observed in AIA 171 Å and STEREO-B/EUVI 171 Å. The black pluses sign mark the positions of the loop. The green lines are slits across the pluses and perpendicular to the loop, the purple line segments along the green slits indicate the width of the loop. The white boxes are the foot-point regions, loop region, and background region, respectively. The white lines that are overlaid on the AIA 171 Å and STEREO-B/EUVI 171 Å images are outlining the magnetic field lines constructed with the PFSS model.

An MHD seismological theory with the cooling effect has been developed and applied to explain the period decrease of loop oscillations[30–32]. Here, we present the analysis of coronal loop oscillations, whose period increase with time during a number of cycles is observed. We focus on a particular observational event and study this phenomenon combining various observational techniques and the theories of coronal loop oscillations. Since the SMS and DEM methods can be used in combination to obtain the plasma parameters in the corona independently, we employ these two tools to diagnose the densities inside and outside of the coronal loop. We compare the results obtained from the two methods, and put forward a physical cause for the yet unreported increase of the period of coronal loop oscillations. Further, we also report the spatial distribution of the amplitude of the fundamental kink oscillation, in order to carry out spatio-magneto-seismology[38–40].

## Results

**Instruments and Observations.**  An M1.2 flare occurred at the eastern limb of the solar disk on October 11, 2013. The flare developed within a mere quarter-an-hour at 07:08 UT and peaked at 07:25 UT. A coronal loop was located southward of the flare beyond the eastern limb as shown in Fig. 1. The coronal loop was observed at the wavelength of 171 Å by AIA, which provides full-disk images of the Sun with a maximum spatial scale of 0.6 arcsec pixel$^{-1}$ and a cadence of 12 s. AIA includes 7 extreme-ultraviolet (EUV) narrow bands (94 Å, 131 Å, 171 Å, 193 Å, 211 Å, 304 Å, and 335 Å), and 3 UV-visible continua. It covers multi-temperatures from 0.06–2 MK[9]. Different bands are advantageous for detecting different structures in the lower corona. For example, most of the coronal loops are observed in 171 Å with a peak response temperature log $T = 5.85$[27,41], and most of the EUV waves and shocks are observed in 193 Å with log $T = 6.2$, 7.25 and 211 Å with log $T = 6.3$[42,43].

The loop we studied is oriented in the north-south direction, crossing the equator of the Sun, and lies beyond the limb of the solar disk as seen in 171 Å by AIA as shown in Fig. 1. However, the loop cannot be identified on the solar disk in 171 Å images observed by STEREO-B, while the active regions where the feet of the loop are rooted can be located in the EUV images of STEREO-B. Combining the observations by AIA and STEREO-B, the geometry of the coronal loop is determined, i.e. it lays in the $x$-$y$ plane, where $z$-axis is assumed to be towards the observer, in the Heliocentric-Cartesian coordinates of SDO. The shape of the loop is asymmetric, namely, the apex of the loop is not at the middle between the two footpoints, but slightly deviates to the south.

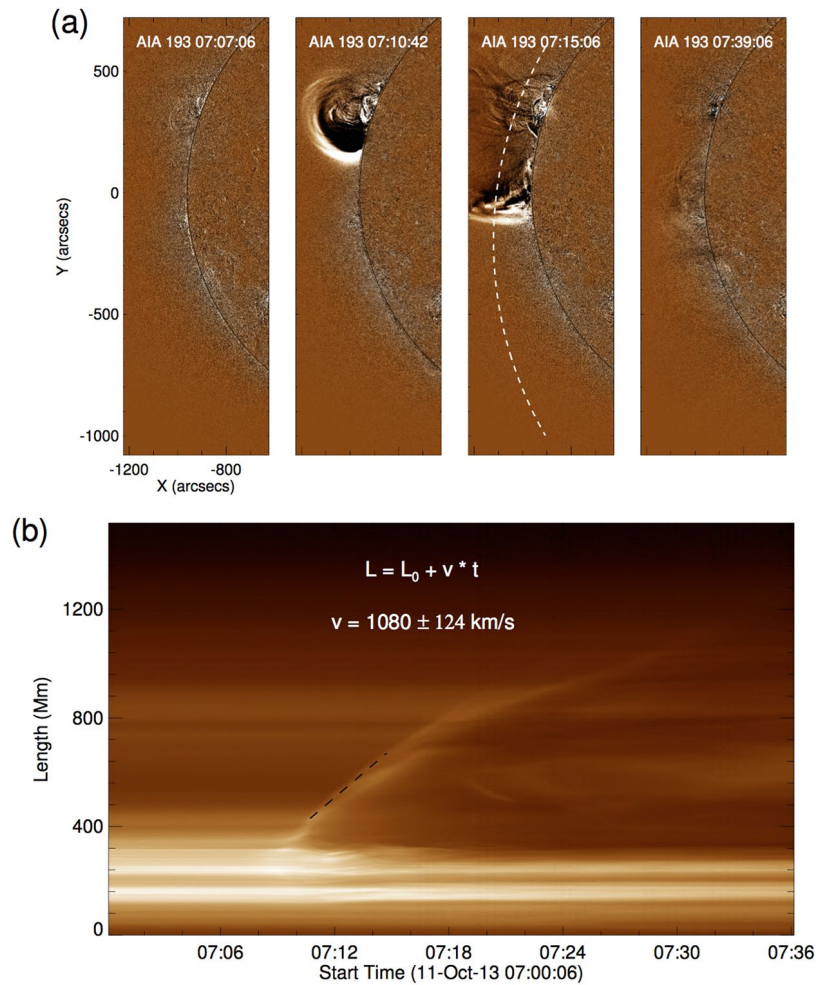

**Figure 2.** Panel (a) displays the running-difference images in 193 Å showing the evolution of the EUV wave that passed through the coronal loop. The white dashed line delineates the propagating direction of the EUV wave from its source to the coronal loop. Panel (b) is the time-distance images of the propagating path of the EUV wave, the black dashed line indicates the positions of the EUV wave front during the propagation.

The geometry of the coronal loop, constructed by using the Potential-Field-Source-Surface (PFSS) model[44,45], is mimicked by the solid white line in Fig. 1. Figure 1a,b show perspectives of AIA and STEREO-B, respectively. The geometric structure of the loop obtained by the PFSS extrapolation is consistent with the excess emission observed by AIA and STEREO-B. The PFSS model assumes potential magnetic field (i.e. $\nabla \times \mathbf{B} = 0$)[44,45], thus fundamentally solving the Laplace equation $\nabla^2 \Psi = 0$, where $\Psi$ represents the scalar potential. Combined with measurements of the field at the photospheric boundary, $\mathbf{B}$ can be deduced from the solution of the Laplace equation governing its potential field using, e.g., spherical coordinate system[46,47].

Next, in Fig. 2, we report the detection of an EUV wave observed by AIA off the east limb, where the EUV wave is prominent at 193 and 211 Å. The wave appeared at 07:09 UT, and it became quickly far too faint to be detectable at 07:21 UT, i.e., not even lasting for 12 minutes. EUV waves are often interpreted as the fast mode magnetosonic waves[48,49]. The loop was in the propagation path of the EUV wave, which was pushed by the wave upon its arrival, and the loop began to oscillate after the wave passing through at about 07:11 UT. The result of this interaction may seem to be a typical coronal loop oscillation triggered by an EUV wave[17,37]. The coronal loop oscillated in the $x$-$y$ plane, identified as a vertically polarised kink mode oscillation[20]. An arc slit is taken along the propagation direction of the EUV wave (Fig. 2a). The associated time-distance diagram is shown in Fig. 2b. We estimate the speed of the EUV wave propagating along the loop from the time-distance diagram, and find the linear speed of the EUV wave in the direction along the loop to be about 1080 km s$^{-1}$.

**Measurement of Parameters of Coronal Loop Oscillation.** To study the loop oscillation, we pinpoint 12 loci along the coronal loop, which are marked as a series of black plus signs in Fig. 1a. Then, we place 12 slits across the points perpendicular to the coronal loop, shown as green lines in Fig. 1a. The time-distance diagrams of the slits from north to south are plotted in Fig. 3. The oscillation of the coronal loop is triggered by the EUV wave propagating from north to south. In about several minutes, the loop becomes fully oscillating after the initial interaction with the EUV wave. The phase of the oscillation profiles of the slits is almost the same when compared to each other, indicating that the detected loop oscillation is a standing wave. The oscillation profiles are most clear in the time-distance diagrams in slits 4, 5, 6 (visualized in Fig. 4), and the amplitudes are decreasing with time. The time intervals between the wave troughs are interpreted as the periods of the oscillations. All three shown profiles have three full periods at least.

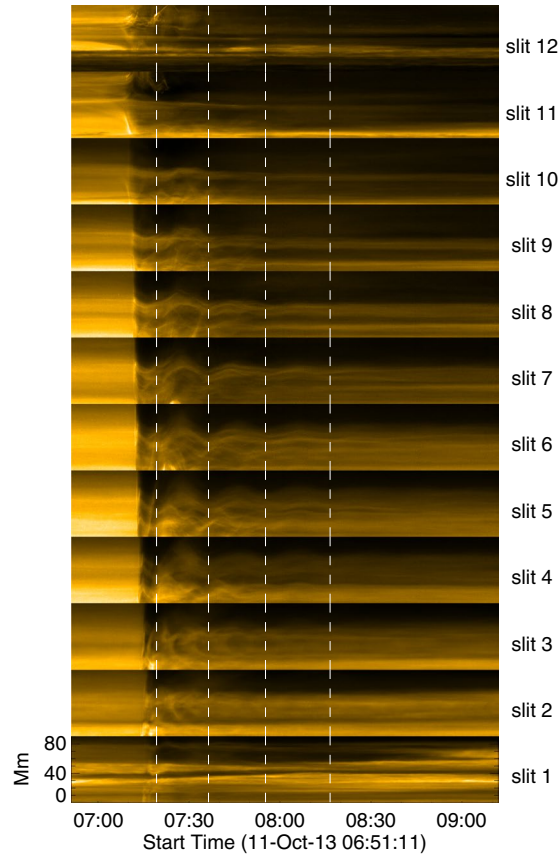

**Figure 3.** The time-distance images of the slits shown in Fig. 1a from north to south. The white dashed lines denote the wave troughs of the intensity profiles in the time-distance images.

Let us now use wavelet analysis to determine the evolution of period of the oscillations. The results are shown in Fig. 4. There is only one peak in the power spectrum for each oscillation profile within the confidence interval. The most powerful periods of the oscillation profiles are at around 1000 s. As shown in Fig. 4, we measure the elapsed time between intervals of the wave troughs during loop oscillation, and find that the time intervals are growing with time. The theory modelling the variation of the period of standing oscillations or propagating waves during coronal loop oscillations has already been developed[30–32]. The decrease of periods is interpreted as a signature of the slow cooling (or evacuation) of the plasma during the oscillation in these studies. Note, however, that we report here for the first time that the period increases during the oscillation, as evidenced in Fig. 4. Now, let us focus on the properties and interpretation of the increase of period during the oscillation.

When the period of oscillations is constant for a magnetic flux tube, the associated amplitude variation of oscillations may be fitted by a trigonometric (e.g. cosine) profile combined with an exponential decay reflecting that only the amplitude may be decreasing (often damped). Nevertheless, here, even the period itself is varying with time. Assuming that the period ($P$) changes with time slowly, $P$ can be written as $P_0 + kt$ empirically, with $k$ being the growth rate. Then, the amplitude is written as:

$$A(t) = A_0 e^{-\frac{t}{\tau}} cos\left(\frac{2\pi t}{P_0 + kt} - \phi\right) + f(t),$$  (1)

where $A_0$ is the initial amplitude, $\tau$ is the damping time, $\phi$ is the initial phase, and $f(t)$ is a function to describe the balance positions of the oscillation. The balance positions of the oscillations are usually fitted by polynomial or spline interpolation[50,51], here, $f(t)$ taken to be a second order polynomial:

$$A_{00} + A_{01}t + A_{02}t^2.$$

From Equation (3) of Morton[32,52], $P$ can also be estimated as: $\sqrt{C + Dt}$. Thus, Equation (1) may be re-cast as:

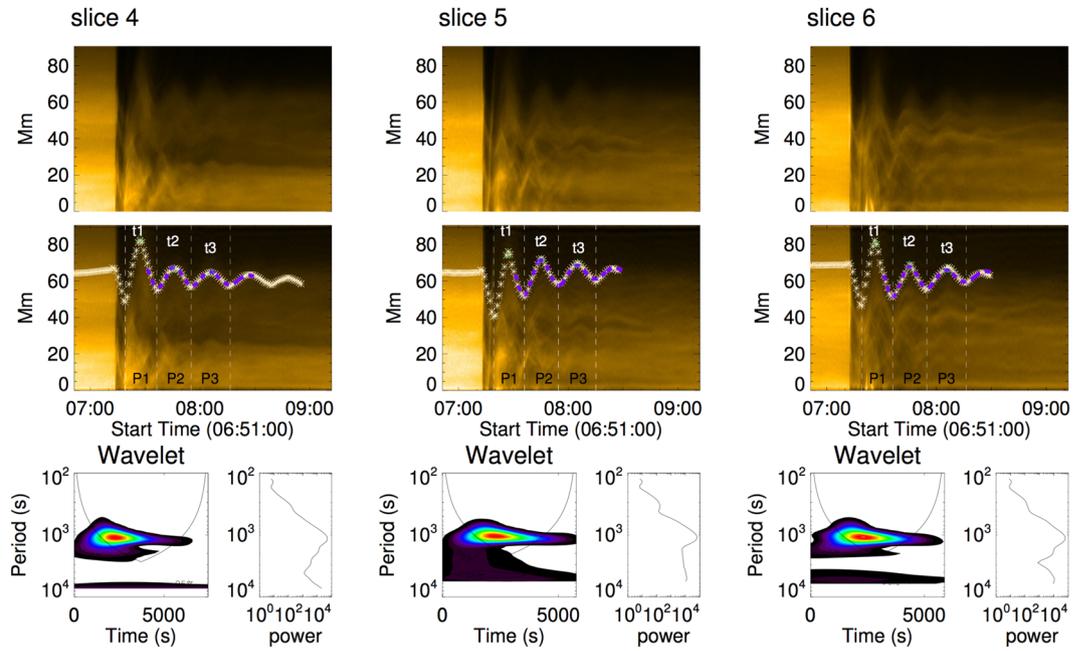

**Figure 4.** Loop oscillation measurements. The upper two rows show the time-distance images of slits 4, 5, and 6. The white asterisks mark the positions that are measured from the time-distance images. The white dashed lines indicate wave trough times of the profiles. The green triangles indicate wave crest moments. The purple dashed lines outline the fitting results using Equation (2). The bottom panels are the power associated spectra obtained by wavelet analysis.

| Parameters | | slit 4 | slit 5 | slit 6 | mean |
|---|---|---|---|---|---|
| Equation (1) | $A_0$ (Mm) | 14.8 | 41.5 | 30.8 | — |
| | $\tau$ (s) | 3231 | 1994 | 2183 | — |
| | $P_0$ (s) | 950 | 1040 | 908 | — |
| | $k$ | 0.0349 | 0.0228 | 0.0438 | — |
| | $\phi$(°) | −22.3704 | 8.20980 | −10.7705 | — |
| | $\chi^2$ | 20.415504 | 15.533685 | 25.385518 | — |
| Equation (2) | $A_0$ (Mm) | 15.0 | 28.4 | 22.0 | — |
| | $\tau$ (s) | 3205 | 2493 | 2696 | — |
| | $C$ (s²) | 828508 | 1045489 | 712675 | — |
| | $D$ (s) | 79.4748 | 53.3842 | 94.8083 | — |
| | $\phi$(°) | −25.3949 | 6.58908 | −16.6835 | — |
| | $\chi^2$ | 19.684437 | 8.1544921 | 18.374332 | — |
| $P1$ (s) | | 1032 | 1044 | 1068 | 1048 |
| $P2$ (s) | | 1128 | 1176 | 1080 | 1128 |
| $P3$ (s) | | 1284 | 1212 | 1296 | 1264 |
| $t1$ (UT) | | 07:27:12 | 07:27:12 | 07:27:12 | — |
| $t2$ (UT) | | 07:46:00 | 07:46:12 | 07:45:00 | — |
| $t3$ (UT) | | 08:06:00 | 08:07:24 | 08:06:36 | — |

**Table 1.** Parameters of the oscillation. $A_0$ (Mm), $\tau$ (s), $P_0$ (s), $k$, $C$(s²), $D$ (s) and $\phi$(°) are the fitting parameters for Equations (1) and (2). $\chi^2$ provides the errors of the curves fitted by Equations (1) and (2), respectively, for the slits 4, 5 and 6. $P1$, $P2$, and $P3$ are the periods for the three time intervals between the troughs. Finally, $t1$, $t2$, and $t3$ are the moments of the three crests.

$$A(t) = A_0 e^{-\frac{t}{\tau}} \cos\left( \frac{2\pi t}{\sqrt{C + Dt}} - \phi \right) + f(t). \tag{2}$$

Equations (1) and (2) are used to fit the oscillation profiles in the time-distance images, as shown in Fig. 4.

In the process of the fitting, we define $t = 0$ at 07:27:12 UT, which is the time when the loop first reaches the crest in Fig. 4. The parameters of the fitting applied to data for slits 4, 5 and 6 are listed in Table 1. The purple curves in Fig. 4 are from Equation (2). The fitted results from Equations (1) and (2) are slightly different. The $\chi^2$ of the fitted curves for the three slits from Equation (1) are larger than those from Equation (2). Empirically, the linear form in Equation (1) is simple to describe the increase of $P$, whereas, Equation (2), where $P$ is estimated as $\sqrt{C + Dt}$, is more reasonable from a physical point of view because radiative cooling of the coronal loop plasma is considered. Combining the difference of $\chi^2$ deduced from Equations (1) and (2), it indicates that the fitted results from Equation (2) are more accurate than those from Equation (1).

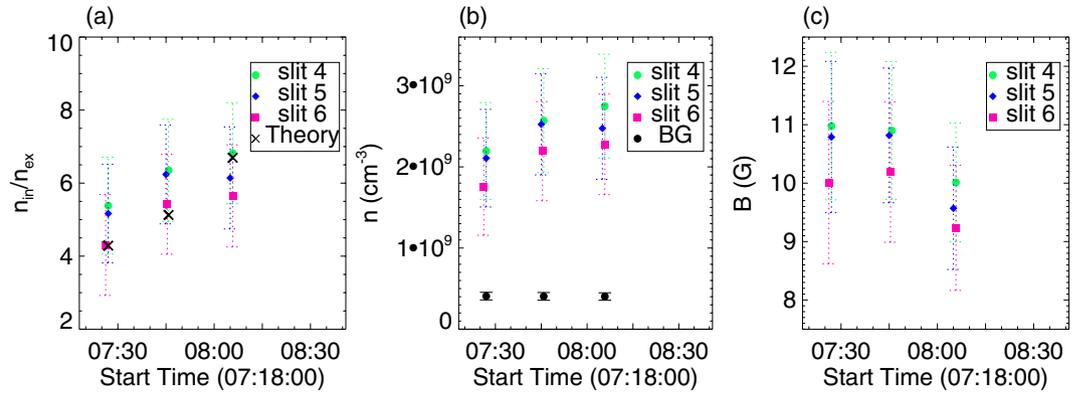

**Figure 5.** Panel (a) displays the evolution of $n_{in}/n_{ex}$. The green circles, purple diamonds, and pink squares mark $n_{in}/n_{ex}$ derived by the DEM method for slits 4, 5 and 6, respectively. The black crosses represent $n_{in}/n_{ex}$ by Equation (4). Panel (b) displays the evolution of $n_{in}$ (green circles, purple diamonds, and pink squares) and $n_{ex}$ (black solid circles). Panel (c) shows the evolution of the magnetic field strength, $B$.

**Period Variation during Oscillation.** The period, $P$, of the coronal kink oscillation is estimated as follows[34]:

$$P = \frac{2L}{v_a} \sqrt{\frac{1 + \rho_{ex}/\rho_{in}}{2}} = \frac{\sqrt{2\mu_0}L}{B}\sqrt{\rho_{ex}(1 + \rho_{in}/\rho_{ex})}, \tag{3}$$

where $L$ is the length, $v_a$ is the Alfvén speed inside the loop, $\rho_{in}$ is the density inside and $\rho_{ex}$ is the density outside the loop, $\mu_0$ is the magnetic permeability, $B$ is the magnetic field strength along the loop.

Equation (3) shows that $P$ is directly proportional to $L$, which is measured to be $392 \pm 9$ Mm at 07:00:00 UT before the start of the loop oscillation, and $377 \pm 7$ Mm at 09:00:00 UT after the loop is back to its initial position. $L$ is slightly longer before the oscillation than after. Since $P$ is directly proportional to $L$, the increase of $P$ cannot be due to the shrinking of $L$.

Now, let us focus on $\rho_{in}$ and $\rho_{ex}$ and their effects on the period $P$ of the oscillation. It has been reported that the dimming regions often appear before a solar eruption[53], and they can be refilled by mass from the footpoint regions of the atmosphere after the ejection[54]. This latter would indicate that $\rho_{ex}$ may be varying during an eruption. Note that $\rho_{in}$ itself may also evolve during solar eruption. We also recall reporting that the densities of the flaring loops increase after the flares[55]. When flares occur and EUV waves pass through coronal loops, the heating and condensation at the footpoints of the coronal loops can lead to the variation of the pressure, temperature and densities of the loops[56]. Thus, we measure the densities inside and outside the loop here. The magnetic field inside and outside the loop can be considered as similar[57], and combining the definitions of Alfvén speeds, the ratio of Alfvén speeds between outside and inside of the loop is equal to the square root of the number density ratio between inside and outside of the loop ($\sqrt{n_{in}/n_{ex}}$). Then, $n_{in}/n_{ex}$ can be calculated theoretically as follows[6]:

$$\frac{n_{in}}{n_{ex}} = \frac{1}{2}\left(v_d\frac{P}{L}\right)^2 - 1, \tag{4}$$

where $v_d$ is speed of the wave that drives the loop oscillation, which is $1080 \pm 124$ km s$^{-1}$ here. The estimates of $n_{in}/n_{ex}$ are shown as black crosses in Fig. 5a, which provide evidence how $n_{in}/n_{ex}$ increases with time. Note that the error of $n_{in}/n_{ex}$ derives from the error of $L$, $v_d$ and $P$, and is determined by using the error transfer formula.

Besides Equation (4), we can also deduce $n_{in}/n_{ex}$ independently, by the DEM method during the loop oscillation. Here, we adopt a model where the EM per unit length is different for different structures existing along the line of sight (LOS)[43]. We modify it to estimate the density ratio between the inside and outside of the coronal loop here (see Methods for the details):

$$\frac{n_{in}}{n_{ex}} = \sqrt{\frac{(l/l_{in})(EM_{lp} - EM_{bg}) + EM_{bg}}{EM_{bg}}}, \tag{5}$$

where $l$ is the effective length of the LOS, and $l_{in}$ is the width of the coronal loop, $EM_{lp}$ and $EM_{bg}$ are the emission measures when the LOS runs across the loop region and background region, respectively. These two variables can be estimated by DEM method from the observations directly. Given that $l \sim \sqrt{H\pi r}$, where $H$ is the scale height, we take $l$ as $4 \times 10^8$ m[58] and $r$ is the heliocentric distance. The loop widths $l_{in}$ are measured by the lengths of the purple line segments along the three slits (shown in Fig. 1a). We take the box outside the loop in Fig. 1a as the background region, and calculate the $EM_{bg}$ at the moments of time of the three wave crests as listed in Table 1. The emission measure, $EM_{lp}$, in the loop regions along three slits is obtained from the observations in the six EUV passbands of AIA at three moments of time when the loop reaches the crest. DEM is also computed for the background region, indicated by the denoted box in Fig. 1a. The DEM curves of the slits, for two foot regions and background region at the three moments, are displayed in Fig. 6. The EM of the foot regions are larger than those of the loop and background regions. The temperature of the slits and background regions are about 1.8 MK at the three moments, i.e. the loop is warm. The temperature of the foot region 2 in Fig. 1 is 5.1, 6.5, and 5.7 MK at the three moments.

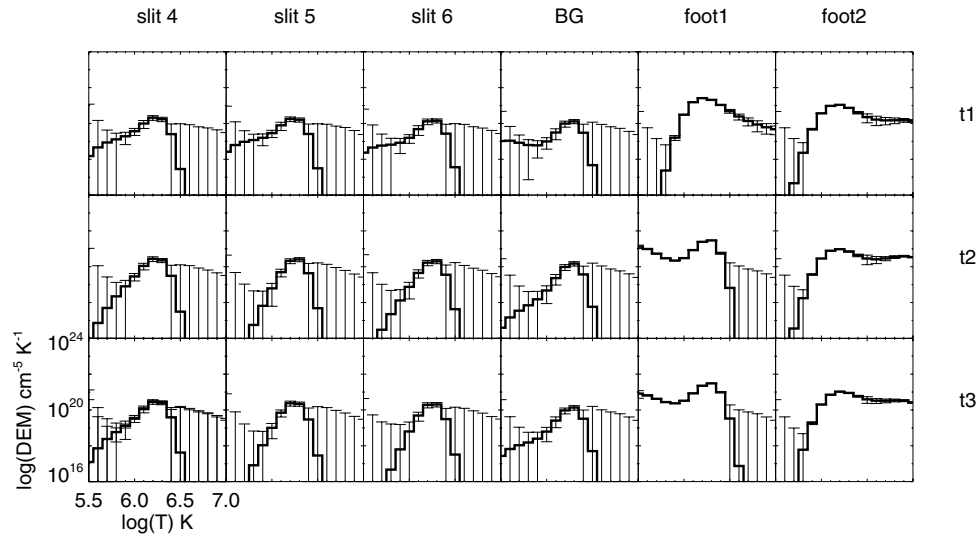

**Figure 6.** DEM results for the loop regions of slits 4, 5, and 6, the background region, foot 1 and foot 2 regions in Fig. 1 at the three moments when the loop reaches the crest in Fig. 4.

Since there are no special structures along the LOS of the background region, the density of the background ($n_{ex}$) is obtained from the DEM method directly. The density ratio, $n_{in}/n_{ex}$ can be calculated by Equation (5), and the density inside the loop ($n_{in}$) is derived consequently. The evolution of $n_{in}/n_{ex}$, $n_{in}$, and $n_{ex}$ for each slit is shown in Fig. 5. The moments of time in Fig. 5 are those of the three wave crests for each slit in Fig. 4. The errors of $n_{in}$ and $n_{ex}$ are generated by the error of the DEM method, the error of $n_{in}/n_{ex}$ are calculated by the error transfer formula of Equation (5). In Fig. 5, $n_{in}$ and $n_{in}/n_{ex}$ increase with time, while $n_{ex}$ is relatively unchanged.

The estimates of $n_{in}/n_{ex}$ obtained from the DEM method are consistent with those obtained by the SMS theory. The consistency manifests that the increase of period is caused by the increase of ($n_{in} + n_{ex}$). Since $n_{ex}$ is relatively constant during the oscillation, the variation of ($n_{in} + n_{ex}$) and $n_{in}/n_{ex}$ is due to the increase in $n_{in}$. This also indicates that the increase of the period of the loop oscillation is primarily caused by the increase of $n_{in}$ during the oscillation.

Further, the increase of $n_{in}$ implies that there may be a mass injection into the loop during the oscillation. Chromospheric evaporation is an effective mechanism that can lead to the density increasing inside the loop, which is often accompanied by flares. Plasma at footpoints of flare loops are heated during the flare and evaporate, which make the pressure inside the loop to become out of equilibrium[59]. Then, coronal loops are filled with dense plasma[56], accompanied by the increase of intensities in EUV and X-Ray wavelengths[55,60]. Figure 7 shows these light curves at the footpoints and of the top loop region in 7 EUV wavelengths of AIA complemented with observations at 195 Å and 304 Å of STEREO-B/EUVI. The intensity values of the light curves in the footpoints are about 10 times larger than those of the loop region, and 20 times larger than those of the background plasma in the EUV wavelengths of AIA. In 195 Å and 304 Å of STEREO-B/EUVI, since the loop is no longer beyond the limb, the background emission of the solar disk mixes with that of the loop region. However, the intensities of the light curves of the footpoints are still larger than those of the loop region. As we have mentioned above, we note that the temperature of the foot region 2 in Fig. 1 is higher than the loop region. The considerable temperature and density differences between the footpoints and loop regions result in pressure differences. Therefore energy is transferred from high temperature regions to low, mass is transferred from high to low density regions, that is to say, the footpoints can supply mass to the loop region. This may be the cause of the increase of $n_{in}$ during the oscillation.

Finally, the strength of the magnetic field, $B$, is also investigated here. Using the densities obtained by the DEM method and the periods we measured from the oscillation profile, the magnetic field strength can be estimated by Equation (3). The evolution of $B$ is shown in Fig. 5c, where $B$ decays with time during the oscillation. It follows from Equation (3), that $B$ is inversely proportional to $P$, the decay of $B$ causes the increase of period during the oscillation here, too. Given that the plasmas-$\beta$ is very low in the corona, the magnetic field was expected to remain constant during the loop oscillation, similar to previous studies[6,36,61,62]. However, $B$ obtained by combining

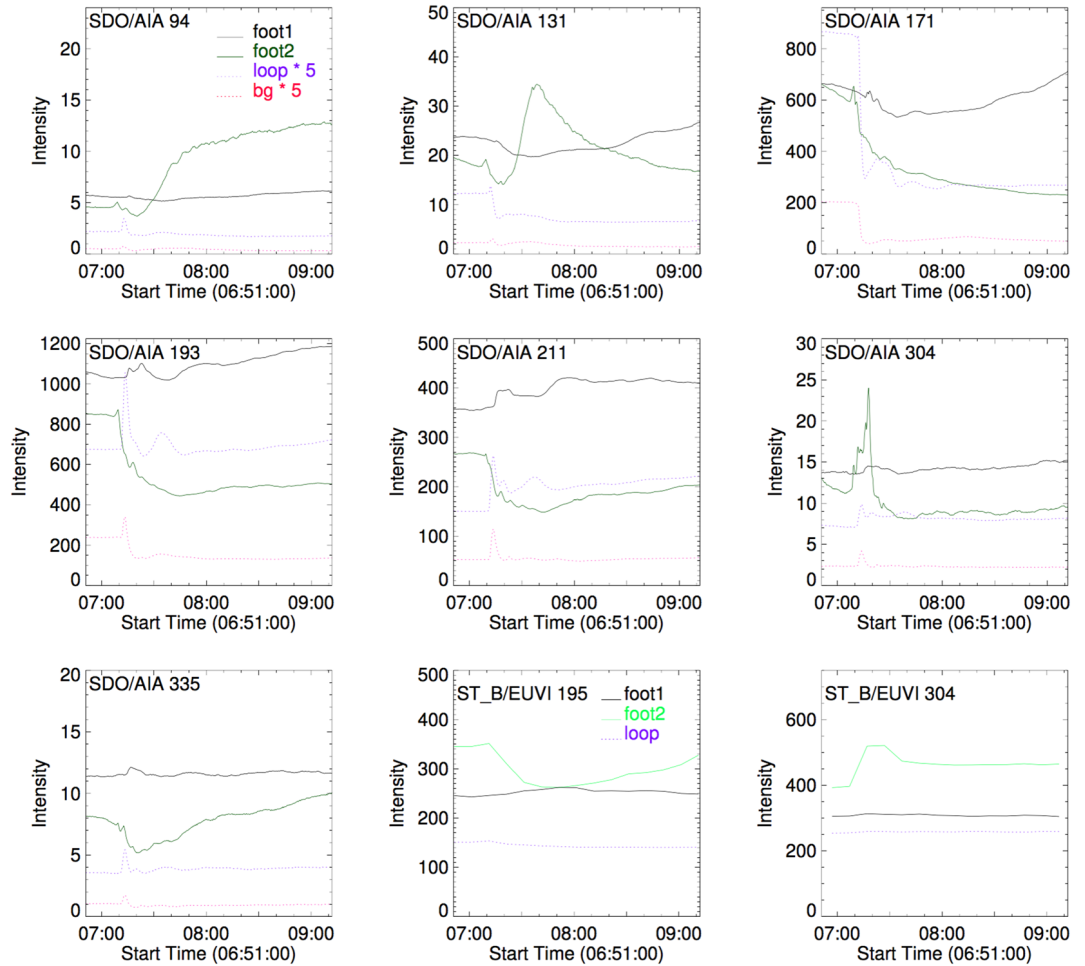

**Figure 7.** The panels show the light curves of the 94 Å, 131 Å, 171 Å, 193 Å, 211 Å, 304 Å, and 335 Å of AIA for the foot, loop, and background regions, and the light curves of the 195 Å and 304 Å, of STEREO-B/EUVI for the foot and loop regions.

SMS theory with the DEM method shows that it is varying during the oscillation. Therefore, it is suggested that the increase of the period during the oscillation is caused by the increase of $n_{in}$ and the decay of $B$ concurrently.

**Amplitude Distribution along the Coronal Loop.**  The height of the magnetic field lines that are obtained from the PFSS model along the coronal loop is indicated by red solid line in Fig. 8. The distribution of the amplitude along the coronal loop is marked as solid diamonds in Fig. 8, which reveals that the distribution is asymmetric. The maximum of the amplitude along the loop is at the apex of the loop, that is slightly shifted to the south footpoint of the loop. Equation 5 is used to estimate $n_{in}$ at the apex of the loop, the values are shown as solid colorful symbols in Fig. 5. Combined with the assumption of stratified density inside coronal loops[63] and the height along the magnetic loop, taking the electron-to-proton abundance is 1.2, the normalized density stratified along the loop is shown as the blue dashed line in Fig. 8. The normalized magnetic field strength, shown in Fig. 8, near the footpoints of the loop is stronger than that at the top. The distribution of magnetic field strength along the loop is also asymmetric, the magnetic strength at the north footpoint is stronger than that at the south footpoint. The minimum of the magnetic field strength is located at around the apex of the loop. The Alfvén speed, $C_{a0}$, is given by $B/\sqrt{\mu_0 \rho_{in}}$, therefore kink speed can be estimate as $C_k = (2/(1 + \rho_{ex}/\rho_{in}))C_{a0}$. Combined with the density stratification and the distribution of the magnetic field strength along the loop, the distribution of the normalized $C_k$ along the loop is shown as the blue dotted line in Fig. 8.

In theory, the amplitude profile of the kink oscillation is a harmonic (e.g. sine) function when the density and magnetic strength is uniform. However, in reality, the density near the footpoints of the loop is larger than at the top, and the magnetic strength near the footpoints of the loop is stronger than that at the top. The amplitude of a loop with uniform magnetic strength and stratified density is larger than that with a uniform density at the same position along the loop (except the footpoints and the loop apex), when the amplitudes at the loop apex are the same for the two cases. Similarly, the amplitude of a loop with uniform density and decreasing magnetic strength is smaller than that with a uniform magnetic strength at the same position along the loop. In Fig. 8, the amplitude of the kink oscillation along the loop is smaller than that of a fitted sine function with the same amplitude at the loop apex. This implies that the amplitude profile along the loop is mainly determined by the distribution of the magnetic field strength along the loop. Only the density stratification in a magnetic loop with a uniform magnetic field

                                                                                8

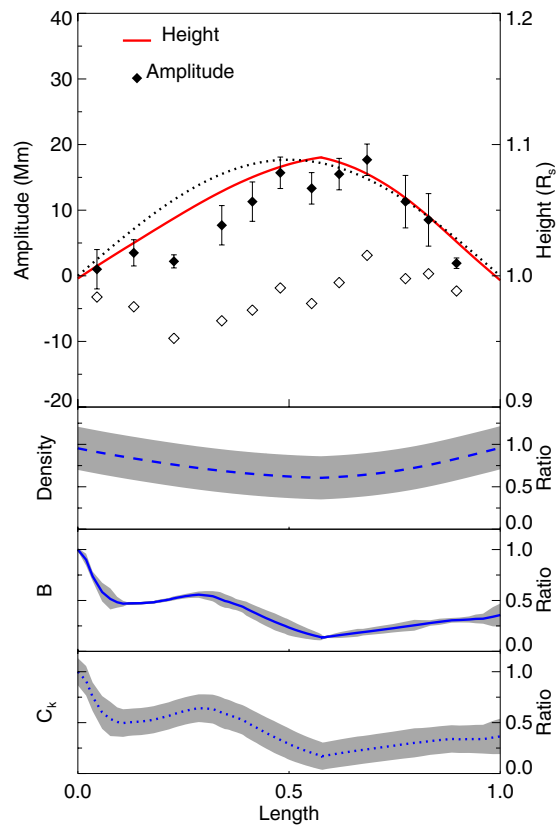

**Figure 8.** Parameters along the coronal loop. The north footpoint of the coronal loop is located at $\overline{L} = 0$, while the south footpoint is located at the normalized $\overline{L} = 1$. The full (black) solid diamonds indicate the maximum amplitude along each slits in Fig. 1 from north to south, the black dotted line marks the fundamental *sine* function that is fitted to the amplitude, and the empty (black) diamonds represent the difference between the fitted amplitude profile and the actually measured amplitude. The red solid line indicates the height of magnetic field line that is obtained from the PFSS model along the coronal loop. The blue dashed line represents the normalized density distribution along the loop. The blue solid line indicates the normalized magnetic field strength along the coronal loop. The blue dotted line indicates the distribution of the kink speed, $C_k$ along the loop. The gray shaded regions are the error bars of density, $B$, and $C_k$, respectively. We note that the density, magnetic field strength, $B$, and $C_k$, are normalized by their own maximum values, respectively.

and constant cross-section would have an opposite effect[38–40,63]. Further, the distribution of the density, magnetic field strength, $B$, and the kink speed, $C_k$, is not only non-uniform, but also asymmetric, as manifested in Fig. 8. The distribution of the amplitude along the loop is also asymmetric.

## Discussion

Based on observed parameters (e.g., period, loop length, and damping time) during coronal loop oscillations, SMS can be applied to carry out diagnostics (e.g., obtaining estimates of density, Alfvén speed, and even magnetic field) of corona loops[34]. Besides SMS, a combination of magnetic field extrapolation and DEM methods are also employed here to diagnose these physical parameters of oscillating loops. The ratio of the inside and outside density of a place can be determined by SMS and the DEM methods independently, and the strength of magnetic field, $B$, can also be determined by SMS and magnetic field extrapolation independently. These multiple techniques can then be combined together to carry out consistency checks and validate the derived estimates.

With the aforementioned techniques, here, we studied a vertically polarised kink oscillations in a coronal loop. The kink oscillations were triggered by an EUV wave passing by the loop. The period of the loop oscillation was measured, and it was found to increase with time. The oscillation profiles that are fitted by Equation (2) are better fit than those by Equation (1), implying that the increase of the period can be attributed to the cooling or heating taking place during the oscillation. The variation of periods during kink oscillations may be due to the expansion of loops with time and the properties inside the loop[64,65]. The period decrease during loop oscillations has already been developed and applied[30–32]. Theoretically, in brief, the period of a kink oscillation is related to the length of the loop, magnetic field strength and densities inside and outside of the loop. The magnetic field around the loop is obtained from PFSS extrapolation, where field lines are reconstructed to match those along the coronal loop observed by AIA and EUVI. The length of the loop is measured, which shows that $L$ decreases only very slightly (3.8%) during the oscillation. Therefore $L$ is excluded to be a dominant reason for the increase of period during the oscillation. The ratio between the inside and outside densities of the coronal loop can be obtained by solar magneto-seismology combined with DEM method independently. Thus, we employ these two methods to estimate $n_{in}/n_{ex}$ during the loop oscillation. $n_{in}/n_{ex}$ obtained by solar magneto-seismology and the DEM method both increased during the oscillation, therefore results are found consistent with each other. The densities inside and outside of the coronal loop cannot be obtained by solar magneto-seismology independently, therefore we use the DEM method to determine them. We found that the increases of $n_{in}/n_{ex}$ and $(n_{in}+n_{ex})$ are by large due to the increase of $n_{in}$ primarily. Besides, using $n_{in}$ and $n_{ex}$ obtained by DEM and Equation (3), we concluded that $B$ decays during the oscillation. Since $P$ is directly proportional to $n_{in}/n_{ex}$ and inversely proportional to $B^{34}$, it can be concluded that the increase of the period of the oscillation is caused by the increase of $n_{in}$ and the decay of $B$ during the oscillation concurrently.

Damping mechanisms such as absorption, leakage, etc., can affect the amplitude[24,28]. However, these damping mechanisms cannot yet be observed directly with the current suit of instrumentation. Here, instead, we propose a natural and simpler mechanism that can be observed and does not require much a priori assumptions.

In theory, when the density and magnetic strength profile along a loop is uniform, the amplitude profile of a kink oscillation is a combination of trigonometric functions, in particular the fundamental mode is a sine function. The amplitude along a magnetically uniform loop with stratified density is larger than that of a loop with a uniform density at the same position, where the magnetic field along the loop is uniform, when the amplitude at the loop apex is the same[38]. On the contrary, the amplitude along a loop with decreased background magnetic strength is smaller than that of a loop with a uniform magnetic strength at the same position, where the density along the loop is uniform, when the amplitude at the loop apex is the same. The effects of density stratification on the amplitude are just opposite to that of magnetic flux tube expansion[63]. Since the amplitude along the loop is smaller than that of a fitted sine function for this event, this indicates that the deviation of the amplitude from the sine function is mainly due to the distribution of the magnetic strength along the coronal loop, rather than the density distribution. This conclusion is also supported by the distribution of the density, magnetic field strength, and the kink speed, which show that the amplitude profile closely mimics the inverse of the distribution of the magnetic field strength and kink speed, rather than the density.

Coronal loops are modelled as ideal 1D structures, and the period of loop oscillation is a parameter that one can occasionally measure. The period of a kink oscillation reflects on the diagnostic properties of the loop. Temporal variation of the period reflects the evolution of the properties of the loop during its oscillations. The variation is usually related to heating or cooling at the footpoint regions, condensation inside the loop, or plasma bulk motion[30–32,66,67]. It is helpful to understand the heating and cooling mechanism of the coronal loops. Besides, the spatial SMS is combined with the observation here, and we find that the amplitude distribution is effected mainly by the magnetic strength along the coronal loop for this event. It implies that we must pay attention to the magnetic strength distribution along the loop when we study the coronal loop oscillation.

## Methods

### DEM method.
The observed flux $F_i$ of each passband is determined by

$$F_i = \int R_i(T) DEM(T) \mathrm{d}T,$$

(6)

where $R_i(T)$ is the temperature response function of each passband $i$, and $DEM(T)$ is the differential plasma EM, dEM, in the differential temperature, dT, in the corona. Observations in six passbands (94 Å, 131 Å, 171 Å, 193 Å, 211 Å, and 335 Å) are used to constrain the distribution of DEM[2]. In this paper, we use the "xrt_dem_iterative2.pro" routine in the SSW package to compute the DEM. This code is originally developed by Weber et al.[68], and then modified to work with AIA[69–72]. This code is widely used to derive DEM from observations[73].

The DEM-weighted average temperature is defined as[73]:

$$T = \frac{\int DEM(T) * T \mathrm{d}T}{\int DEM(T) \mathrm{d}T}.$$

(7)

We can apply Equation (7) to calculate the temperature of coronal plasma.

The total EM is:

$$EM = \int DEM(T) \mathrm{d}T = \int n^2 \mathrm{d}l,$$

(8)

where $n$ is plasma density.

### The density ratio between the inside and outside loop.
When different structures (e.g. a coronal loop) pass through the LOS, the EM cannot be considered as uniform. We exhibit a model for this situation here[43].

From Fig. 9, there are two LOS, where LOS1 passes through a loop, and there are no loops across LOS2. For the background region, the EM per unit length ($em_{ex}$) along the LOS can be considered as uniform[37,43,73]. Since LOS1 and LOS2 are close to each other, $em_{ex}$ along LOS1 and LOS2 can be considered as the same. The EM along LOS2 is expressed as:

$$EM_{bg} = em_{ex} l,$$

(9)

where $em_{ex}$ denotes the EM per unit length of the background region, $l$ is the total effective length of the LOS.

When the coronal loop runs across the LOS, the total EM consists of two parts, one from the loop and the other from the background:

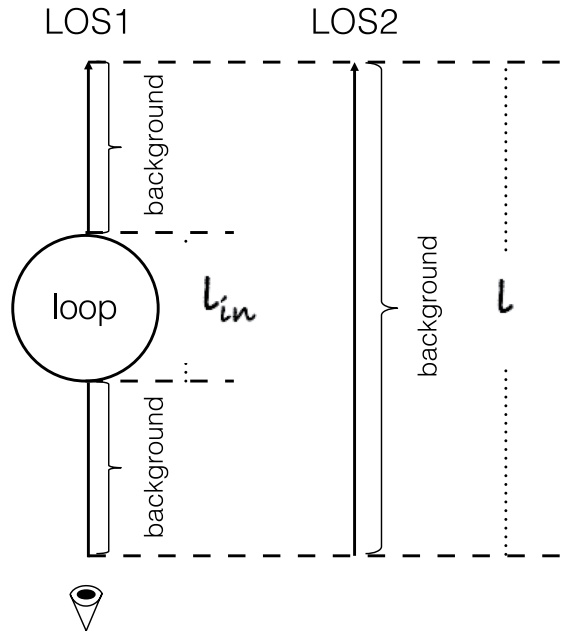

**Figure 9.** Schematic showing the effective depth contributing to the EM along the LOS, modified from Fig. 9 in Su *et al.*[43]. LOS1 passes through a loop along the LOS, while LOS2 only passes through the background region.

$$EM_{lp} = \frac{l - l_{in}}{l} EM_{bg} + EM_{in} = em_{ex}(l - l_{in}) + em_{in}l_{in},$$ (10)

$em_{in}$ is the EM per unit length inside the coronal loop, $l_{in}$ is the width of the coronal loop.

From Equation (8) the density ratio between inside and outside loop $(n_{in}/n_{ex})$ can be written as $\sqrt{em_{in}/em_{ex}}$, thus, Equations (5) can be obtained from Equations (9) and (10).

## Acknowledgements

We thank the SDO team for providing the EUV images, the STEREO team for 3D observations, and Schrijver C. and DeRosa M. for providing PFSS code. We are grateful to Li B. and Chen P.F. for helpful discussions and valuable suggestions about the theory of coronal loop oscillations. This work is supported by Jiangsu NSF (Grants No. BK20171108, BK20161095, BK20161618), KLSA201611, NSFC (11533005, 11573072, 11603077, 11773016, 11733003, 11733061, 11790302 (11790300)), Jiangsu 333 Project (BRA2017359) and Laboratory No. 2010DP173032. Erdélyi R. is grateful to Science and Technology Facilities Council (STFC, Grant No. ST/M000826/1) UK, the Royal Society (UK) and acknowledges the support received from the CAS President's International Fellowship Initiative, Grant No. 2016VMA045. Erdélyi R. and Guo Y. are also grateful for the support received from the 2016 Sheffield International Mobility Scheme and from the Mathematics and Statistics Research Centre (MSRC), University of Sheffield for their international collaboration and bilateral visits that enhanced this research.

## Author Contributions

Su W. processed the data and analyzed the results. Guo Y., Erdélyi R. and Ding M.D. provided detailed analysis and theoretical interpretation of the results, Ning Z.J. proposed the research, Cheng X. provided the DEM method, and Tan B.L. offered crucial suggestions.

## Additional Information